\documentclass[aps,12pts,pre,amsmath,amssymb,preprint,showkeys,showpacs]{revtex4}
\usepackage{graphicx}
\usepackage{bm}
\usepackage{mathptmx}
\usepackage{booktabs}
\usepackage{multirow}
\usepackage{hhline}
\usepackage{dcolumn}
\usepackage{subfigure}
\usepackage{amsmath}
\usepackage{float}
\usepackage{epsfig}
\setcitestyle{super}
\begin{document}
\title{An improved d-band model of the catalytic activity of magnetic transition metal surfaces}
\author{Satadeep Bhattacharjee$^{1}$, Umesh V. Waghmare$^{2}$ and S. C. Lee$^{1,3}$} 
\email{leesc@kist.re.kr,seungcheol.lee@ikst.res.in}
\thanks{Corresponding author}
\address{$^{1}$Indo-Korea Science and Technology Center (IKST), Bangalore, India \\
         $^{2}$Jawaharlal Nehru Centre for Advanced Scientific Research(JNCASR), Bangalore, India \\
$^{3}$Electronic Materials Research Center, Korea Institute of Science $\&$ Tech, Korea}
         
\begin{abstract} {\bf The d-band center model of Hammer and N{\o}rskov is widely used in understanding and predicting catalytic activity on transition metal (TM) surfaces. Here, we demonstrate that this model is inadequate for capturing the complete catalytic activity of the magnetically polarized TM surfaces and propose its generalization. We validate the generalized model through comparison of adsorption energies of the NH$_3$ molecule on the surfaces of 3d TMs (V,~Cr,~Mn,~Fe,~Co,~Ni,~Cu and Zn) determined with spin-polarized density functional theory (DFT)-based methods with the predictions of our model. Compared to the conventional d-band model, where the nature of the metal-adsorbate interaction is entirely determined through the energy and the occupation of the d-band center, we emphasize that for the surfaces with high spin polarization, the metal-adsorbate system can be stabilized through a competition of the spin-dependent metal-adsorbate interactions.}
\end{abstract}

\keywords{d-band center model, spin catalyst, two-state reactivity}
\pacs{82.65.My, 82.20.Pm, 82.30.Lp, 82.65.Jv}

\maketitle

\section*{I\lowercase{ntroduction}}
Due to the low abundance, toxicity-related issues and high cost of 4d and 5d metals, in recent years researchers have turned to developing catalysts using cheap and abundant 3d transition metals (TMs) and their alloys or oxides~\cite{3d1,3d2,cheap}. The catalytic reactions in these materials can also be manipulated using spin, in addition to the usual parameters such as size, strain, and electrode potential. The role of magnetism in heterogeneous catalysis is the subject of a recent study~\cite{metiu,groot,pick,beat,water}. It was demonstrated by Behler \textit{et al.}\citep{sheffler} that on metal surfaces such as Al (111), spin selection leads to a low sticking probability of O$_2$ molecules with a triplet spin state. Recently, Melander \textit{et al.}~\cite{crossover} showed that the reactivity of metal surfaces is dependent on their magnetic states. Using first-principles methods, they noted that in the case of adsorption of H$_2$ on a ferromagnetic Fe surface, there is an asymmetry in the Fe-H$_2$ interaction for majority and minority spin channels. Such asymmetric interaction results in weaker hydrogen-metal binding for a ferromagnetic Fe surface than for an antiferromagnetic Fe surface. In the ferromagnetic case, only spin minority electrons take part in the bond formation, while on the antiferromagnetic surface, the bond formation is accomplished through both the minority and majority spin electrons.

Such notable results obtained either from the first-principles simulations or experiments require a simple theoretical model to interpret. ~The majority of the first-principles theoretical studies are focused on the understanding the nature of interaction between the adsorbate and the d-electrons of the TM surface~\cite{TM0,TM0-1,TM1,TM2,TM3,TM4}. The most widely employed model invoked to understand the role of the d-electrons is the so-called \textit{d-band center model}~\cite{d-band1,d-band2,d-band3,d-band4,d-band5} of Hammer and N{\o}rskov, developed more than a decade ago. This simple yet highly celebrated model of chemisorption is again based on the concepts of other models of chemisorption such as (1) the Newns-Anderson model~\cite{Newns,Anderson} and (2) the effective medium theory~\cite{e1,e2,e3}. The former is a more general description of the interaction of the adsorbate state with the continuous band of valence states of the metal, while the latter relates the adsorption energy to the local electron density and the change in one-electron states of the surface.


In the d-band model, the band of d-states participating in the interaction is approximated with a single state at energy $\varepsilon_d$, known as the \textit{center of the d-band}. Such a model can be thought of as a \textit{narrow d-band limit} of the Newns-Anderson model. According to this model, the variation in the adsorption energy from one TM surface to another correlates the upward shift of this d-band center with respect to the Fermi energy. A stronger upward shift indicates the possibility of the formation of a larger number of empty anti-bonding states, leading to a stronger binding energy. The upward shift of the d-band center can therefore be treated as a \textit{descriptor} of the catalysis. Hammer-N{\o}rskov model successfully explains both the experimental and the first-principles theoretical results for different ligands/molecules on a variety of TM surfaces~\cite{val1,val2,val3}.
\par
However, there are few studies on the adsorption of molecules on metal surfaces with high spin polarization. Moreover, if an adsorbate itself has a considerable magnetic dipole moment, it will have a strong magnetic interaction with the surface. Therefore, the validity of the Hammer-N{\o}rskov model for molecular adsorption on surfaces with large spin polarization is not obvious. The d-band center model predicts a uniform decrease~(increase) of the adsorption energy of a given molecule from one TM surface to another where the number of d-electrons increases~(decreases). An exception to the prediction of the d-band center model occurs for OH adsorption on Pt and Pd skin alloy systems~\cite{exc}. However, such exceptions are typically related to the large electronegativity of the adsorbate and the substrate having a nearly full d-band.

In the present study, we demonstrate the limitations of the conventional d-band center model via a simple case study: the adsorption of non-magnetic molecules such as NH$_3$ on 3d TM surfaces. The reaction of NH$_3$ on TM surfaces is important due to its relevance in controlling the corrosion of steel and iron surfaces. We show that for a better comparison with the results obtained from the spin-polarized DFT-based methods, the conventional d-band center model has to be extended by considering two band centers, one each for the spin majority and the spin minority electrons of the system. Such a model would be useful in designing chemical reactions that can be controlled through spin arrangement of the catalytic surface or by an external magnetic field.
\section*{A\lowercase{desorption on 3d}-TM \lowercase{surfaces; why do we need a spin-polarized d-band center model}?}
Here, we examine the applicability of the conventional d-band center model to a simpler problem: adsorption of non-magnetic molecules on spin-polarized metal surfaces. From a comparison of the adsorption energies of an NH$_3$ molecule on 3d TMs obtained from spin-polarized and spin-unpolarized calculations, we find a significant effect of spin polarization on adsorption. The adsorption energies of the molecule on magnetic surfaces are smaller for the spin-polarized calculations~(see Fig.~1). This simple fact also suggests that the d-band center model, which relies on a non-spin-polarized (or spin-averaged) description of the surface electrons, has to be expanded to incorporate spin polarization effects.

To understand the trend in catalytic activity across TMs, one should consider the spin polarization of the metal surface in addition to the number of d-electrons. In Fig.~\ref{Schamatic}, we schematically compare the d-band center of a metallic surface with and without spin polarization. When spin polarization is considered in a calculation, it is appropriate to consider two d-band centers, one for the spin up states $\varepsilon_{d\uparrow}$ and the other for the spin down states $\varepsilon_{d\downarrow}$. These are shifted in opposite directions in energy relative to the unpolarized d-band center, $\varepsilon_d$. $\varepsilon_{d\uparrow}$ is shifted downwards,  while the $\varepsilon_{d\downarrow}$ is shifted upwards with respect to $\varepsilon_d$. If we consider that these two centers interact with the adsorbate level, we should obtain two sets of bonding and anti-bonding orbitals that are higher and lower in energy with respect to the unpolarized bonding and anti-bonding levels. The possibility of obtaining a non-linear dependence of the adsorption energy with the number of d-electrons originates from the fact that the contributions to the adsorption energy from two such band centers can \textit{compete} with each other. Naturally, when the degree of the spin polarization is smaller, the two d-band centers are close to each other, and their activity is similar. However, when the spin polarization is higher, the two band centers are shifted significantly in opposite directions. If we consider the interaction with an adsorbate possessing multiple levels, among which the occupied ones are closer to the metal band centers than for the minority spin, there are more unoccupied metal-adsorbate anti-bonding states giving rise to strong attractive interactions, while there are more occupied metal-adsorbate states for the majority spin electrons, resulting in strong repulsion. Therefore, the minority spin d-bands bind more strongly to the adsorbate, while the binding with majority spin states is weaker.  This phenomenon results in large changes in the adsorption energies of Mn and Fe as shown in Fig.~1 for spin-polarized and non-spin-polarized cases. 
\section*{T\lowercase{wo-centered d-band model}}
In this section, we generalize the d-band model but still follow the approach used by Hammer and N{\o}rskov. Let us consider the interaction of the adsorbate states with the metal states using a basis set with a minimum number of states, $\{\psi_{ai\sigma},\psi_{d\sigma}\}$, where $\psi_{ai\sigma}$ is the i$^{th}$ adsorbate state with spin $\sigma$ and $\psi_{d\sigma}(\sigma=
\uparrow,\downarrow)$ are two hypothetical discrete states representing metal states with two spins. The adsorption energy can be expressed as follows (see the supplemental material):
\begin{equation}
\Delta E_d=-\sum_{\sigma,\sigma',i}\frac{f_{\sigma}{V^{da^i}_{\sigma,\sigma'}}^2}{|\varepsilon_{ai\sigma'}^*-\varepsilon_{d\sigma}|}
-\sum_{\sigma,\sigma',j}(1-f_\sigma)\frac{{V^{da^j}_{\sigma,\sigma'}}^2}{|\varepsilon_{d\sigma}-\varepsilon_{aj\sigma'}|}+
\sum_{\sigma,\sigma',i}f_\sigma \alpha{V^{da^i}_{\sigma,\sigma'}}^2+
\sum_{\sigma,\sigma',j} (1+f_\sigma)\alpha{V^{da^j}_{\sigma,\sigma'}}^2
\label{E1}
\end{equation}
For simplicity, we have assumed that all the adsorbate states are sigma-type orbitals. $V^{da^k}_{\sigma,\sigma'}$ are the matrix elements of the coupling between the TM d-state with the k$^{th}$ adsorbate state, $\varepsilon_{ai\sigma'}^*$ is the energy of the i$^{th}$ unoccupied adsorbate state with spin $\sigma'$ and $\varepsilon_{aj\sigma'}$ is the energy of the j$^{th}$ occupied adsorbate state. The two d-band centers for the majority and minority spins are respectively $\varepsilon_{d\uparrow}$ and $\varepsilon_{d\downarrow}$. The first term in the above equation is the energy gain due to the interaction of the unfilled adsorbate state with the metal states. The second term describes the interaction of the metal d-states with the filled adsorbate states. The first term always describes an attractive interaction, while the second term has both attractive and repulsive components. Here, $f_\sigma$ is the fractional filling of the metal state with spin $\sigma$. The last two terms in Eqn.~\eqref{E1} are due to the orthogonalization of the adsorbate state and TM d-states and are always repulsive. The parameter $\alpha$ is adjustable and has units of eV$^{-1}$. The third term is due to the orthogonalization of the empty adsorbate states on the metal d-states, while the fourth term represents the orthogonalization of the filled adsorbate states on the metal states. 
\par 
When there is more than one adsorbate state, with some filled and some empty, to understand how the net attractive and repulsive interactions compete with each other in a realistic situation, we consider the case of an NH$_3$ molecule on the TM surface. In this case, the adsorbate is non-magnetic, and we assume that the interaction parameter $V^{da}_{\sigma,\sigma'}$ is spin-independent and constant for a particular metal surface. We express the various energy contributions to the molecule and d-electron interaction as follows: 
\begin{equation}
\Delta E_d=-\sum_{\sigma}\sum_{i=1}^N \frac{f_\sigma V^2}{|\varepsilon_{ai}^*-\varepsilon_{d\sigma}|}
-\sum_{\sigma}\sum_{j=1}^M(1-f_\sigma)\frac{V^2}{|\varepsilon_{d\sigma}-\varepsilon_{aj}|}+N
\sum_{\sigma,\sigma'}f_\sigma \alpha V^2+M
\sum_{\sigma,\sigma'} (1+f_\sigma)\alpha V^2
\label{E2}
\end{equation}
where $N$ and $M$ are respectively the number of unoccupied and occupied adsorbate orbitals.
\subsection*{S\lowercase{pin-dependent attractive and repulsive surface-adsorbate interaction}}
Eqn.~\eqref{E2} describes a simplified model for adsorption energy of a non-magnetic molecule interacting with the TM surface. The states with energy $\varepsilon^*_{ai}$ and $\varepsilon_{aj}$ are respectively empty antibonding and filled bonding molecular states. Competition and cooperation between the different spin channels during the process of adsorption are evident as we split the energy given by Eqn.~\eqref{E2} into attractive and repulsive parts. The first term in Eqn.~\eqref{E2} is always attractive for arbitrary filling of the d-states for both the spins, while the second term can be written as the sum of attractive and repulsive contributions. The attractive component is as follows:
\begin{equation}
E_{attractive}=-\sum_{\sigma,i}\frac{f_\sigma V^2}{|\varepsilon_{ai}^* -\varepsilon_{d\sigma'}|}
-\sum_{\sigma,j}\frac{V^2}{|\varepsilon_{d\sigma}-\varepsilon_{aj}|}
\label{E3}
\end{equation}
The first term of Eqn.\eqref{E3} gives the gain in energy due to the empty adsorbate levels interacting with the d-band centers, while the second term is the energy gain due to the bonding orbitals formed between the filled adsorbate states and the d-band centers. The energy due to the repulsive interaction between the molecule and the metal surface is given as follows:
\begin{equation}
E_{repulsive}=\sum_{\sigma,j,} \frac{f_\sigma V^2}{|\varepsilon_{d\sigma}-\varepsilon_{aj}|}+ N\sum_{\sigma} f_\sigma\alpha V^2 +M\sum_{\sigma} (1+f_\sigma)\alpha V^2
\label{E4}
\end{equation}
The first term of Eqn.~\eqref{E4} is the energy of the antibonding orbitals, which promotes destabilization of the adsorbate on the metal surface, while the last two terms result from the orthogonalization of the metal and adsorbate states, as already mentioned.
\section*{R\lowercase{esults and discussions}}
In this section, we quantify the energy contributions mentioned in Eqns.~(\ref{E2},\ref{E3}, and \ref{E4}). We have calculated the matrix elements V for different TM surfaces using Harrisons $ d^{\frac{7}{2}}$ rule \cite{Harrison}:
\begin{equation}
V^{da}_{\pi({\bf \sigma})}=\eta^{da}_{\pi({\bf \sigma})}\frac{\hbar^2r_d^{3/2}}{md^{7/2}},
\label{E5}
\end{equation} 
where $\eta^{da}_{\pi({\bf \sigma})}=1.36(-2.95)$, $\frac{\hbar^2}{m}=7.62 eV \AA^2$ are constants.
The characteristic length $r_d$ of the d-orbitals of different TM atoms is taken from Ref.~\cite{Harrison}. The bond-lengths $d$ were taken from our DFT calculations. Because no $\pi$-type molecular orbital is involved, we have considered $V^{da}_{\bf \sigma}$ only (note that here, ${\bf \sigma}$ indicates the type of adsorbate orbital, not the spin index, as in Eqn.~\eqref{E2}). To calculate the attractive and repulsive contributions in Eqns.~(\ref{E3} and \ref{E4}) of an NH$_3$ molecule on different TM surfaces, we considered four discrete energy levels of the NH$_3$ molecule (obtained from the DFT calculations) in a symmetric manner, two from the HOMO region and two from the LUMO region, (see Fig.~\ref{NH3-dos}, where the density of states (DOS) of the NH$_3$ molecule is shown). The DOS exhibits five distinct peaks at the energies $\tilde{\varepsilon_{aj}}$=-15.4 eV, -5.5~eV,~and -0.5~eV and $\tilde{\varepsilon^*_{ai}}$=~4.4 eV,~and 6.4 eV, respectively. Among these peaks, the peak at -5.5 eV corresponds to the doubly degenerate N-H bonding molecular orbital with 1e symmetry, while the peak at -0.5 is due to the molecular orbital with 3a1 symmetry representing the lone pair. The peaks at 4.4 eV and 6.4 eV are the anti-bonding molecular states with symmetries 4a1 and 2e, respectively. In our calculation of the chemisorption energy, we have not considered the level at -15.4 eV since it is energetically too far from both the majority spin and the minority spin d-band centers for all the TMs.

The adsorption energies are calculated from Eqn.~\eqref{E2}, where the renormalized adsorbate levels $\varepsilon_{aj}$ and $\varepsilon^*_{ai}$ are due to the interaction with sp electrons of the metal. These levels are obtained using the Newns-Anderson model~\cite{Newns,Anderson}~(see Fig.~\ref{Interaction}, where the renormalized levels are shown alongside the d-projected DOS of the Fe (110) surface).~The corresponding renormalized DOS of the NH$_3$ molecule is as follows:  
\begin{equation}
D_{NH_3}(E)=\frac{1}{\pi}\sum_j\frac{\Delta(E)}{\left(E-\tilde{\varepsilon_j}-\Lambda(E)\right)^2
+\Delta(E)^2}
\label{ads}
\end{equation}
where $\Delta (E)=\pi \tilde{V}^2D_{sp}(E)$ is the chemisorption function. $\tilde{V}$ describes the adsorbate-metal coupling for the sp electrons, and $D_{sp}(E)$ is the DOS of the metals sp electrons. $\Lambda(E)=\frac{1}{\pi}\int \frac{\Delta(E')}{E'-E} dE'$ is the Kramers-Kr{\"o}nig transformation of $\Delta(E)$. The renormalized adsorbate levels $\varepsilon_{a_j}$ are calculated from the values of $E$ for which the lines described by $y=E-\tilde{\varepsilon_j}$ cross $\Lambda(E)$.
 
In the actual calculation of $\Delta (E)$, we assume a semi-elliptical sp band centered at the Fermi energy, with the bandwidth obtained from our DFT calculation.

In Table 1, in the 2nd and 3rd columns, we show calculated d-band centers for the majority spin and the minority spin for TM surfaces in the 3d series. The fourth column gives the attractive contribution to the metal-ligand interaction, while the fifth column gives the repulsive part of the metal-ligand interaction. Table 1 shows that for V,~Cr,~Cu and Zn,~ $\varepsilon_{d\uparrow} \simeq \varepsilon_{d\downarrow}$ which is not the case for Mn,~Fe,~Co, and Ni, for which $\varepsilon_{d\uparrow} < \varepsilon_{d\downarrow}$. The 6th and 7th columns give the magnitude of the spin-dependent attractive interaction, while the 8th and 9th columns give the magnitude of the spin-dependent repulsive interaction. It is evident from the table that for V,~Cr,~Cu and Zn, the energies for the attractive interaction are the same for both the majority and minority spins. Additionally, as expected, the energies for the repulsive interaction are the same for both the spins. In contrast, for Mn,~Fe,~Co, and Ni (see columns 8 and 9 of Table 1), the attractive interaction has a larger magnitude for the minority spin, while the repulsive interaction has a larger energy for the majority spin. 

In the case of NH$_3$, the strongest molecule-TM interaction is through the filled lone pair~\cite{TM3,lonepair2}. For spin-polarized surfaces, most of the repulsive interaction is produced by the majority spin electrons, mainly because $(1+f_\uparrow)\alpha V^2 > (1+f_\downarrow)\alpha V^2$ since $f_\uparrow > f_\downarrow$.
\par
In Fig. \ref{Comparison}, we show the adsorption energies obtained from the spin-polarized DFT calculations alongside the $\Delta E_d$ calculated from our model (left panel) using Eqns. [\ref{E3},\ref{E4}]. For comparison, we also show $\Delta E_d$ calculated from the Hammer-N{\o}rskov model (right panel). The d-band centers in this case were obtained from the spin-unpolarized DFT calculations. It is evident from Fig.~\ref{Comparison} that our model is more consistent with the trend of the adsorption energies representing the DFT calculation. This better fit arises because the spin-dependent competing metal-adsorbate interaction (which is important for Mn,~Fe,~Co, and~Ni) is absent in the Hammer-N{\o}rskov model.
\par
Instead of the spin-averaged d-band center, $<\varepsilon>=\sum_\sigma\frac{f_\sigma\varepsilon_{d\sigma}}{\sum_\sigma f_\sigma}$, we propose that the adsorption energies obtained from the spin-polarized DFT calculation (or from the experiments) can be correlated with the following descriptor:
\begin{equation}
\varepsilon^{eff}=\sum_\sigma\frac{f_\sigma\varepsilon_{d\sigma}}{\sum_\sigma f_\sigma}-(\varepsilon_{d\downarrow}-\varepsilon_{d\uparrow})\mu
\label{des}
\end{equation}
where $\mu=\frac{f_\uparrow-f_\downarrow}{f_\uparrow+f_\downarrow}$ is the reduced fractional moment.
The first term is the usual spin-averaged d-band center, while the second term is a shift depending on the spin polarization of the surface. The second term is non-zero only for the surfaces with a non-zero magnetic moment. The role of this term is to push the effective d-band center to lower energy and thus capture the effect of the spin polarization in reducing the adsorption energy.~For f$_\uparrow$=f$_\downarrow$ and $\varepsilon_{d\uparrow}=\varepsilon_{d\downarrow}=\varepsilon_d$, $\varepsilon^{eff}=\varepsilon_d$, the descriptor for the usual d-band center model.
In Fig.~\ref{Descriptor}, we plot $\varepsilon^{eff}$ with the adsorption energies obtained through spin-polarized DFT calculations and show the spin-averaged d-band center for comparison.

\section*{G\lowercase{eneral relationship between chemisorption energy and d-band centers in spin-polarized systems}}
The variation of the chemisorption energy from one metal surface to another as predicted in the conventional d-band center model \cite{linic} is as follows:
\begin{equation}
\delta \Delta E_d=\left(\frac{\partial\Delta E_d}{\partial\varepsilon_d}\right)_V \delta \varepsilon_d+\left(\frac{\partial\Delta E_d}{\partial V^2}\right)_{\varepsilon_d} \delta V^2
=\gamma\delta \varepsilon_d+\nu\delta V^2
\label{96-1}
\end{equation}
The first term in Eqn.~\eqref{96-1} corresponds to the covalent interaction between the metal and the adsorbate, while the second term corresponds to the Pauli repulsion due to orthogonalization~\cite{variation} of TM and adsorbate states.
$\gamma=\left(\frac{\partial\Delta E_d}{\partial\varepsilon_d}\right)_V<0$, while  $\nu=\left(\frac{\partial\Delta E_d}{\partial V^2}\right)_{\varepsilon_d} > 0$. 
$\varepsilon_d$ is the d-band center that is obtained either from a non-spin-polarized calculation or through spin averaging, $\varepsilon_d=\sum_\sigma\frac{f_\sigma\varepsilon_{d\sigma}}{\sum_\sigma f_\sigma}$. Ignoring the second term, we obtain the following:
\begin{equation}
\delta \Delta E_d=\left(\frac{\partial\Delta E_d}{\partial\varepsilon_d}\right)_V \delta\varepsilon_d
=\gamma\delta \varepsilon_d
\label{96-2}
\end{equation}
Eqn.~\eqref{96-2}~represents the central result of the conventional d-band center model~\cite{d-band1,variation}, i.e., a positive shift in $\delta \varepsilon_d$ implies an increase in the chemisorption energy, while a negative shift in $\delta \varepsilon_d$ decreases the chemisorption energy.

The variation of the chemisorption energy and the d-band center has the following relationship from our spin-generalized model from Eqn.~\eqref{E2}:
\begin{align}
\delta \Delta E_d=&\left(\frac{\partial\Delta E_d}{\partial\varepsilon_{d\uparrow}}\right)_V\delta \varepsilon_{d\uparrow}
+\left(\frac{\partial\Delta E_d}{\partial \varepsilon_{d\downarrow}}\right)_V\delta \varepsilon_{d\downarrow}
+
\left(\frac{\partial\Delta E_d}{\partial V^2}\right)_{\varepsilon_{d\uparrow}}\delta V^2
+
\left(\frac{\partial\Delta E_d}{\partial V^2}\right)_{\varepsilon_{d\downarrow}}\delta V^2 \nonumber\\
=&\sum_{\sigma}\left(\gamma_\sigma \delta\varepsilon_{d\sigma}+\nu_\sigma \delta V^2\right)
\label{vary}
\end{align}
where $\gamma_\sigma =\left(\frac{\partial\Delta E_d}{\partial\varepsilon_{d\sigma}}\right)_V$ and 
$\nu_\sigma=\left(\frac{\partial\Delta E_d}{\partial V^2}\right)_{\varepsilon_{d\sigma}}$.\\
The form of Eqn.~\eqref{vary} suggests a decrease in the chemisorption energy as we move from a minimally spin-polarized surface to a highly spin-polarized one, since if we consider $\delta \varepsilon_{d\uparrow}$ to be positive, $\delta \varepsilon_{d\downarrow}$ should be negative, and the first two terms in Eqn.~\eqref{vary} will compete. The change of the chemical reactivity due to the antiferromagnetic-to-ferromagnetic crossover~\cite{crossover} can also be understood in terms of Eqn.~\eqref{vary}. For antiferromagnets, there are two spin sub-lattices, which we label A and B. If we consider the simplest case, in which both of the sub-lattices are composed of the same metal, we have 
\begin{equation}
\begin{split}
\varepsilon^A_{d\uparrow}=\varepsilon^B_{d\downarrow}\\
\varepsilon^A_{d\downarrow}=\varepsilon^B_{d\uparrow}
\end{split}
\label{sub}
\end{equation}
From Eqn.~\ref{sub}, the stronger coupling of an adsorbate to the minority spin channel of the sub-lattice A implies a strong coupling to the majority spin channel of the sub-lattice B. This coupling can lead a change of site preference, even in a mono-component antiferromagnetic material.
\section*{S\lowercase{toner criterion and chemisorption}}
The formation of local moment on the i$^{th}$ site of a TM surface is governed by the local Stoner criterion, $D_i(E_F)I > 1$, where $D_i(E_F)$ is the DOS of the d-electrons on i$^{th}$ site at the Fermi energy and I is the Stoner integral. Since strong chemisorption pushes a large number of states from the region near the Fermi energy to lower energies (due to bond formation with the adsorbate), it therefore disturbs the Stoner criterion locally. Thus, these two effects, viz, chemisorption and the Stoner criterion, oppose each other. The former leads to an increase in the kinetic energy, while the latter promotes a smaller kinetic energy so that the magnetism is retained. It is therefore expected that the spin-polarized surfaces would show lower activity than the non-spin-polarized surfaces.

\section*{O\lowercase{utlook}}
It should be noted here that this approach of considering multiple d-band centers can be further extended to study catalytic reactions involving TM oxides, which will help us design inexpensive catalysts~\cite{cheap}. The d-bands of such systems are usually not continuous and contain multiple subbands, mainly due to the crystal field effect. The number and the arrangement of such subbands depends on the symmetry of the crystal field. If the system is magnetic, these subbands further split into minority and majority spin subbands. A reliable description of the catalytic activity of such systems can be obtained only from a model with a Hamiltonian of
$\mathcal{H}=\sum_{a,\sigma}\varepsilon_{a\sigma} n_{a\sigma}+\sum_{i,\sigma}\varepsilon_{di,\sigma}n_{d,i}+
\left( \sum_{a,\sigma}V c^{\dag}_{a\sigma}c_{di\sigma}+H.C \right)$, which describes the interaction between the adsorbate levels $\varepsilon_{a\sigma}$ with a set $\{i\}$ of spin-dependent d-band centers $\{\varepsilon_{di\sigma}\}$ with occupations $n_{di\sigma}$. $c^{\dag}_{a\sigma},~c_{a\sigma}$ are respectively the creation and annihilation operators for the adsorbate states, while $c^{\dag}_{di\sigma},~c_{di\sigma}$ are the corresponding operators for the d-states. For perovskites with an ABO$_3$ structure, $i\in t_{2g},e_g$. Additionally, using the present approach allows one to investigate how to activate the reactions that are forbidden due to conservation of the spin angular momentum~\cite{groot}, by choosing a catalyst material with appropriate spin polarization. Although so-called \textit{two-state reactivity} has already been the subject of a case study of organometallic complex catalysts~\cite{tsr}, the concept was not discussed rigorously for heterogeneous catalysts, most importantly using the concept of d-band centers (narrow d-band limit). There are catalytic reactions in which both the reactants and the products are non-magnetic, but the reaction intermediates can be magnetic, and the rate-determining steps can depend on the spin exchange between the adsorbate and the surface. A more complete analysis along this direction is a subject of future studies.
\section*{M\lowercase{ethods}}
The adsorption energies and the spin-dependent band centers are calculated from first principles. These first-principles calculations are performed within the framework of DFT with the Perdew-Burke Ernzerhof exchange correlation energy functional~\cite{pbe} based on a generalized gradient approximation. We used a projector augmented wave method as implemented in Vienna \textit{ab initio} simulation package (VASP)~\cite{vasp}. The surfaces were modelled as slabs of 4x4 in-plane unit cells and four atomic layers containing 64 atoms. Kohn-Sham wave functions of the valence electrons were expanded in a plane wave basis with an energy cut-off value of 450 eV. Brillouin zone sampling was conducted using a Monkhorst Pack grid of 3x3x1 k-points. Ionic relaxation was performed using the conjugate-gradient method until forces were reduced to within 0.02 eV/Angstrom for the non-constrained atoms. A vacuum of 10~$\AA$ was included. In all cases, we considered close-packed structures of TM surfaces. We considered ferromagnetic (011) surfaces of V,~Cr,~Mn and Fe, the (0001) surface of Co, and the (111) surfaces of fcc Ni, Cu and Zn. The dipole corrections were applied along the directions perpendicular to the metal surface to eliminate the unwanted electric fields arising from the asymmetry of the simulation cell. The structural relaxations were performed for NH$_3$ and only the top two layers of the TM surface. The bottom two layers were fixed to their bulk experimental values. The adsorption energy was calculated from the following relation:
\begin{equation}
E_{ad}=E_{S+A}-(E_S+E_A),
\label{ads}
\end{equation}
where $E_{S+A}$ is the energy of the surface plus the adsorbate and $E_S$ and $E_A$ are the energy of the surface and adsorbate, respectively. We used Eqn.~\eqref{ads} to calculate the adsorption energies of NH$_3$ on different TM surfaces with and without spin polarization. The d-band centers of both the majority spins and the minority spins were calculated from the first moment as given by~\cite{d-band3},
\begin{equation}
\varepsilon_{d\sigma}=\frac{\int_{-\infty}^\infty E D_{d\sigma}(E-E_F)dE}{\int_{-\infty}^\infty \ D_{d\sigma}(E-E_F)dE},
\end{equation}  
where $ D_{d\sigma}(E)$ is the DOS projected on the d-states of the TM for spin $\sigma$ and $E_F$ is the Fermi energy of the system. The spin-dependent fractional occupations are considered as follows: $f_\sigma=\frac{\int_{-\infty}^{E_F} D_{d\sigma}(E)dE }{5}$. These band centers and occupations were used as inputs for Eqns.~(2, 3, and 4).

\section*{A\lowercase{uthor contributions}}
S.B. and S.C.L conceived the idea. S.B performed the numerical and analytical calculations, wrote the paper. S.C.L and U.U.W provided valuable input.
\section*{C\lowercase{ompeting financial interests}}
The authors declare no competing financial interests.
\section*{A\lowercase{cknowledgement}}
UVW acknowledges funding from the Indo-Korea Institute of Science and Technology and support from a J.C. Bose National Fellowship (Dept. of Science and Technology, Govt. of India). This work was supported by the Convergence Agenda Program (CAP) of the Korea Research Council of Fundamental Science and Technology (KRCF).            

\newpage
\begin{table}
\begin{tabular}{|c|c|c|c|c|c|c|c|c|}
\hline
TM  & $\varepsilon_{d\uparrow}$ & $\varepsilon_{d\downarrow}$  & $E_{attractive}$ & $E_{repulsive}$& ($E_{attractive}$)$_{\sigma=\uparrow}$ & ($E_{attractive}$)$_{\sigma=\downarrow}$ & ($E_{repulsive}$)$_{\sigma=\uparrow}$ & ($E_{repulsive}$)$_{\sigma=\downarrow}$ \\
\hline
\hline
V&-0.89 & -0.89 & -1.22 & 0.42 & -0.61 & -0.61 & 0.21 & 0.21\\
Cr&-1.00 & -1.00 & -4.31 & 1.75 & -2.15 & -2.15 & 0.88 & 0.88\\
Mn&-1.38 & -0.29 & -4.20 & 1.50 & -1.05 & -3.15 & 0.78 & 0.72\\
Fe&-1.86 & 0.48 & -1.33 & 0.95 & -0.66 & -0.67 & 0.63 & 0.32\\
Co&-1.93 & 0.28 & -0.22 & 0.16 & -0.10 & -0.12 & 0.09 & 0.07\\
Ni&-1.65 & -1.01 & -1.48 & 1.08 & -0.56 & -0.92 & 0.52 & 0.56\\
Cu&-2.09 & -2.09 & -3.91 & 3.93 & -1.95 & -1.95 & 1.96 & 1.96\\
Zn&-3.88 & -3.88 & -17.90 & 17.63 & -8.95 & -8.95 & 8.81 & 8.81\\

\hline
\end{tabular}
\caption{Calculated d-band centers (in eV) for both the majority and the minority spins for the different TM surfaces. The attractive energy and the repulsive energy due to the molecule-surface interaction and their corresponding values for different spins are also tabulated.}
\end{table}
\clearpage
\newpage
\begin{figure}[t]
\begin{center}
  \includegraphics[width=0.8\columnwidth]{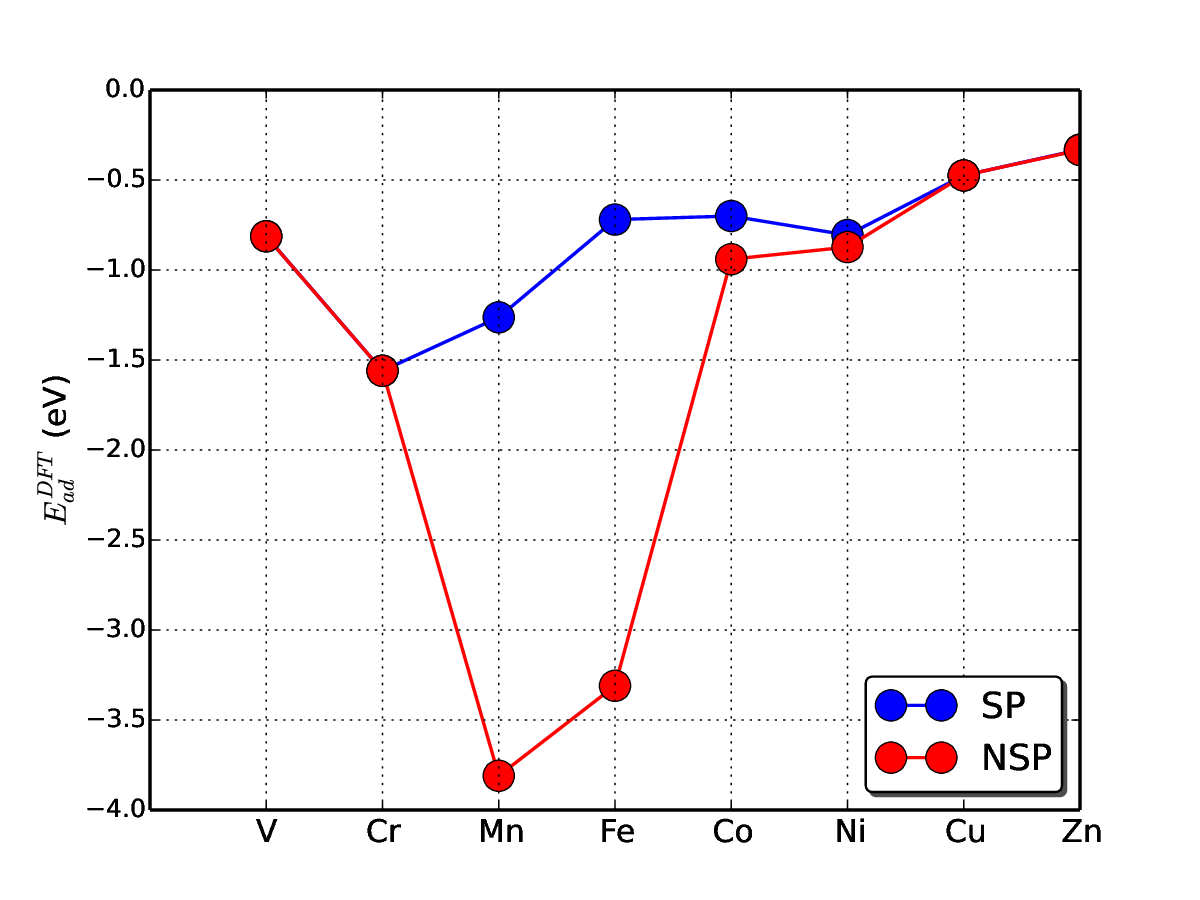}
   \end{center}
\vspace{20mm}
\caption{(Color online) Adsorption energy with the number of d-electrons for spin-polarized~(SP) and non-spin-polarized calculations~(NSP).}
\label{DFT-ad}
\end{figure}
\newpage
\begin{figure}[t]
\begin{center}
\vspace{10mm}
  \includegraphics[width=0.85\columnwidth]{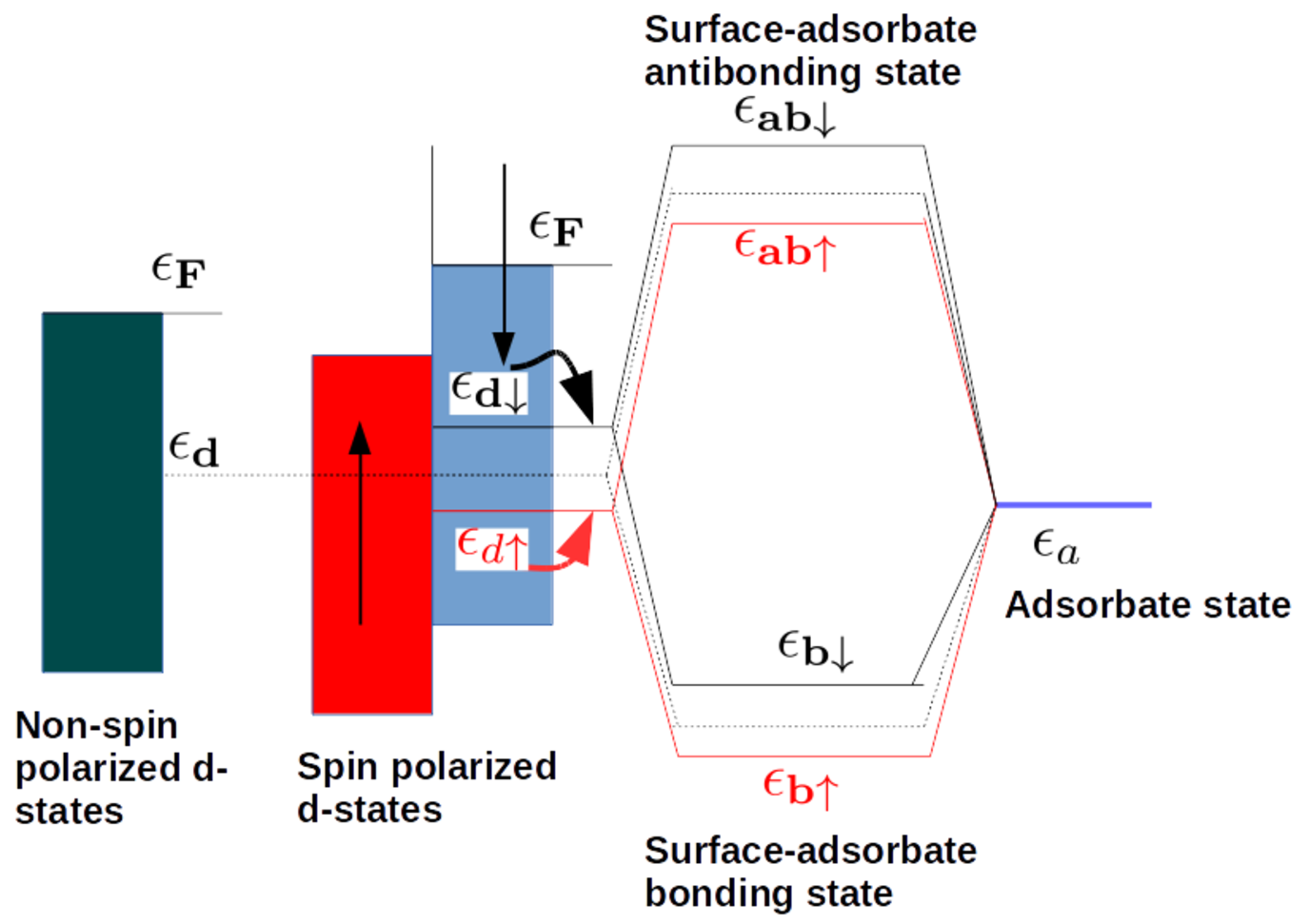}
   \end{center}
\caption{(color online)~Schematic representation of the comparison of the coupling of an adsorbate level $\varepsilon_{a}$ with the metal d-states characterized by a single d-band center~(dotted line), $\varepsilon_{d}$ for the non-spin-polarized case and two d-band centers, and $\varepsilon_{d\uparrow}$ and $\varepsilon_{d\downarrow}$ for the spin-polarized case. $\varepsilon_{b\uparrow(\downarrow)}$ and $\varepsilon_{ab\uparrow(\downarrow)}$ are respectively the metal-adsorbate bonding and anti-bonding energy levels for the majority (minority) spins. E$_F$ is the Fermi energy.}
\label{Schamatic}
\end{figure}
\newpage
\begin{figure}[t]
\begin{center}
\vspace{10mm}
  \includegraphics[width=0.85\columnwidth]{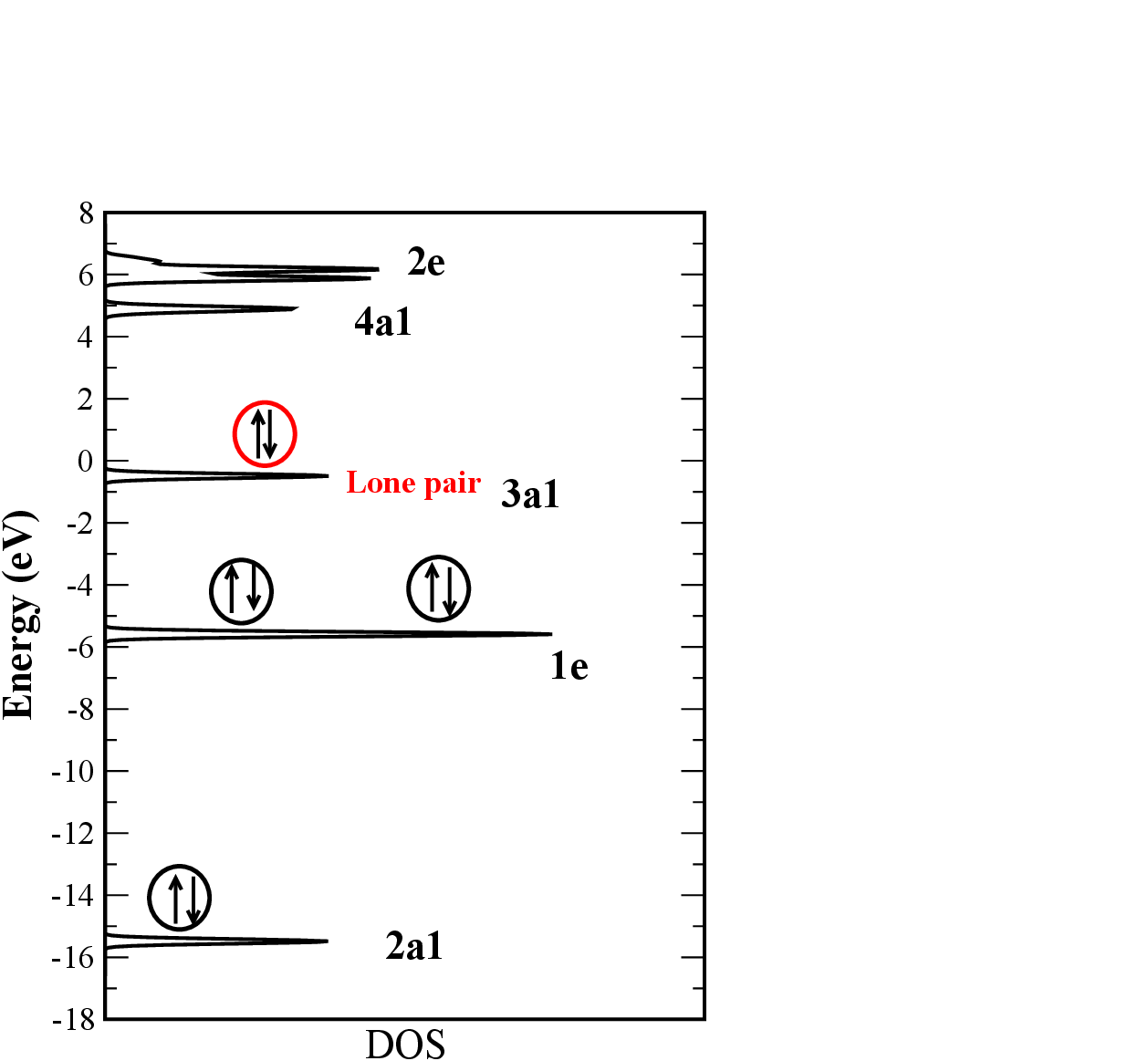}
   \end{center}
\caption{(Color online) The DOS of the NH$_3$ molecule in its gas phase. The filled molecular levels are shown.}
\label{NH3-dos}
\end{figure}
\newpage
\begin{figure}[t]
\begin{center}
\vspace{10mm}
  \includegraphics[width=0.85\columnwidth]{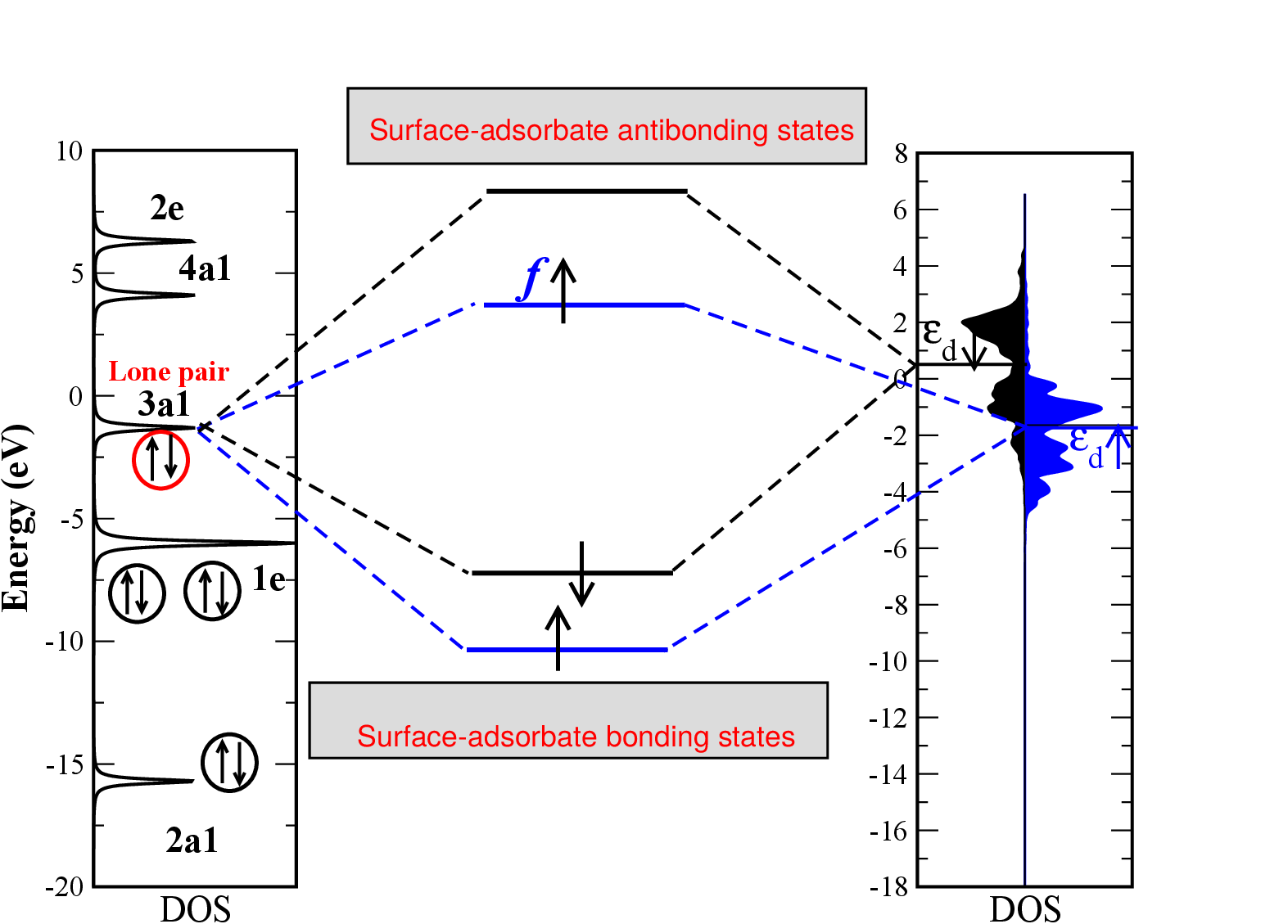}
   \end{center}
\caption{(Color online) The (renormalized) DOS of the NH$_3$ molecule (left) and of the Fe (110) surface (right) as obtained from spin-polarized DFT calculations. The results are shown with respect to the Fermi energy.
As an illustration, we show the interaction of the NH$_3$ lone-pair level (at -0.5 eV) with the two d-band centers ($\varepsilon_{d\uparrow}$ and $\varepsilon_{d\downarrow}$) of the Fe (110) surface. It can be easily understood that the lone-pair-$\varepsilon_{d\downarrow}$ interaction is attractive (the metal-adsorbate bonding and anti-bonding states are shown as black) since $\varepsilon_{d\downarrow}$ is unoccupied, while the lone-pair-$\varepsilon_{d\uparrow}$ interaction (the metal-adsorbate bonding and anti-bonding states are shown as blue in this case) has a repulsive contribution as well because $\varepsilon_{d\uparrow}$ has an occupation of f$_\uparrow$. The magnitude of the bonding-anti-bonding split $\sim \frac{1}{\vert \varepsilon_{d\uparrow(\downarrow)}-\varepsilon_{lone-pair} \vert}$ is larger for minority spin.}
\label{Interaction}
\end{figure}
\newpage
\begin{figure}[t]
\begin{center}
\vspace{10mm}
  \includegraphics[width=0.85\columnwidth]{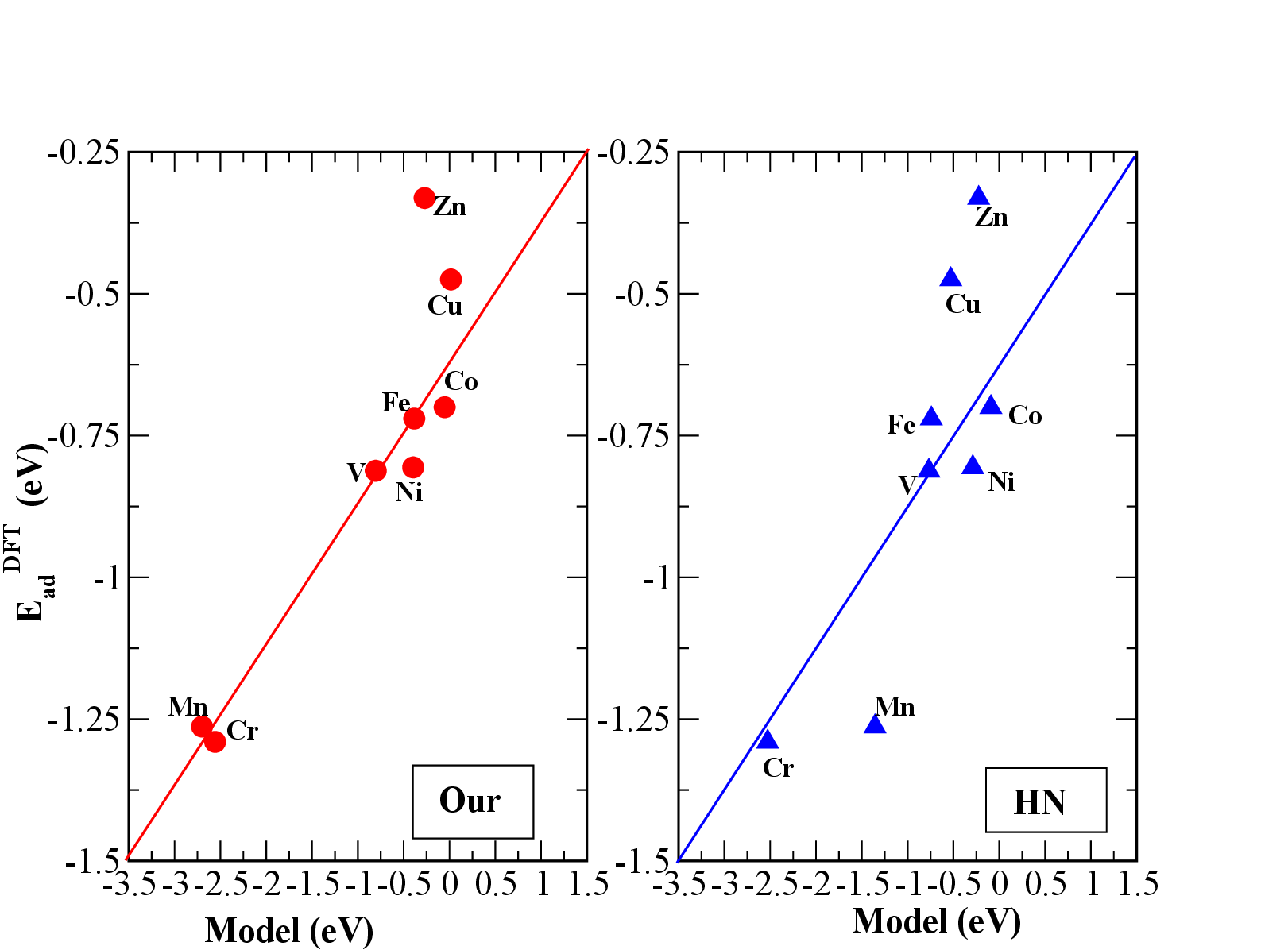}
   \end{center}
\caption{(Color online) The calculated values of  $\Delta E_d$ for different TMs as obtained from Eqn.~2~are compared with the adsorption energies $E_{ad}^{DFT}$ obtained from the spin-polarized DFT calculations (Our). The adsorption energies calculated using the Hammer-N{\o}rskov 
(HN) model are also shown.}
\label{Comparison}
\end{figure}
\newpage
\begin{figure}[t]
\begin{center}
\vspace{10mm}
  \includegraphics[width=0.85\columnwidth]{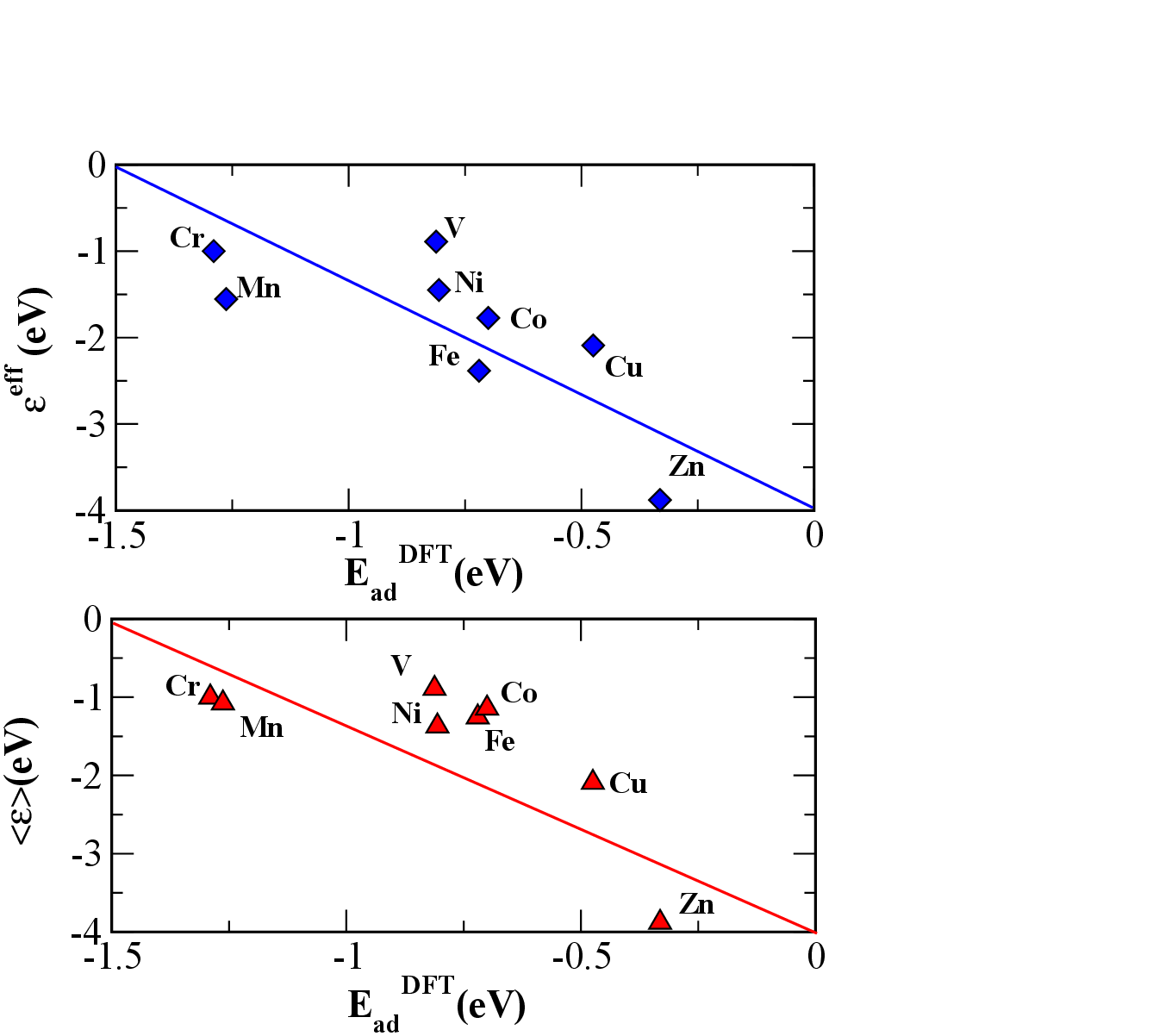}
   \end{center}
\caption{(Color online) Adsorption energies of the NH$_3$ molecule ($E_{ad}$) on different 3d TM surfaces obtained from spin-polarized DFT calculations are shown as a function of the proposed descriptor (top panel) and the spin-averaged d-band center (bottom panel).}
\label{Descriptor}
\end{figure}
%
%
%
%
%
%
\end{document}



\title{Supplemental material of \\``An improved d-band model of the catalytic activity of magnetic transition metal surfaces''}

\author{S. Bhattacharjee$^{1}$, U. V. Waghmare$^{2}$ and S. C. Lee$^{1,3}$}
\address{$^{1}$Indo-Korea Science and Technology Center (IKST), Bangalore, India \\
         $^{2}$Jawaharlal Nehru Centre for Advanced Scientific Research(JNCASR), Bangalore, India \\
         $^{3}$Electronic Materials Research Center, Korea Institute of Science $\&$ Tech, Korea}


\maketitle

\vspace{2mm}\def\b#1{\boldsymbol{#1}}

\large{\bf Derivation of two-centered d band model }\\
\vskip 0.1cm
Let us consider the simplest case, an adsorbate with a single molecular orbital is interacting with a 
transition metal (TM) surface represented by its d-band centers $\varepsilon_{d\sigma}$. $\sigma$ is the spin index($\sigma=\uparrow,\downarrow$). Supposing that the metal and the adsorbate sub-systems are characterized by the wave functions respectively
$\Psi_{d\sigma}$ and $\Psi_{a\sigma}$, i,e 
\begin{equation}
{\cal H}_d\Psi_{d\sigma}=\varepsilon_{d\sigma}\Psi_{d\sigma}
\end{equation}
\\ 
and 
\begin{equation}
{\cal H}_a\Psi_{a\sigma}=\varepsilon_{a\sigma}\Psi_{a\sigma}
\end{equation}
when they are not coupled. $\varepsilon_{a\sigma}$ is the energy of the adsorbate state 
with spin $\sigma$.
The wavefunction of the interacting system , ${\cal H}_{tot}
={\cal H}_d+{\cal H}_a+{\cal H}_{da}$, ( where ${\cal H}_{da}|\Psi_{d\sigma}>=V|\Psi_{a\sigma}>$
describes the mixing between a and d states)
can be written in terms of linear combination of atomic orbitals (LCAO) approach,
\begin{equation}
\Psi_{da\sigma}=C_{d\sigma}\Psi_{d\sigma}+C_{a\sigma}\Psi_{a\sigma}
\end{equation}
the energy of the interacting system is given by,
\begin{equation}
E=\frac{<\Psi_{da\sigma}|{\cal H}_{tot}|\Psi_{da\sigma}>}{<\Psi_{da\sigma}|\Psi_{da\sigma}>}
\end{equation}
Due to the interaction there will be spin-dependent bonding and anti-bonding orbitals which can be obtained 
by setting:\\
$$\frac{\partial E}{\partial C_{d\sigma}}=0$$ and $$\frac{\partial E}{\partial C_{a\sigma}}=0$$
\begin{equation}
\begin{aligned}
\varepsilon_{b\uparrow}=\frac{\varepsilon_{d\uparrow}+\varepsilon_a}{2}-VS-\sqrt{\frac{4V^2+(\varepsilon_{d\uparrow}-\varepsilon_a)^2}{2}},\\
\varepsilon_{ab\uparrow}=\frac{\varepsilon_{d\uparrow}+\varepsilon_a}{2}-VS+\sqrt{\frac{4V^2+(\varepsilon_{d\uparrow}-\varepsilon_a)^2}{2}}, \\
\varepsilon_{b\downarrow}=\frac{\varepsilon_{d\downarrow}+\varepsilon_a}{2}-VS-\sqrt{\frac{4V^2+(\varepsilon_{d\downarrow}-\varepsilon_a)^2}{2}},\\
\varepsilon_{ab\downarrow}=\frac{\varepsilon_{d\downarrow}+\varepsilon_a}{2}-VS+\sqrt{\frac{4V^2+(\varepsilon_{d\downarrow}-\varepsilon_a)^2}{2}}
\end{aligned}
\end{equation}
Where we have considered $\varepsilon_{a\uparrow}=\varepsilon_{a\downarrow}=\varepsilon_a$. $V=<\Psi_{d\sigma}
|{\cal H}_{da}|\Psi_{a\sigma}>$, are independent of the spin. The subscript 'b' refers the bonding state, while 'ab' refers the anti-bonding state. The situation is shown in the Fig.1 of the manuscript. The overlap integral is defined by $S=<\Psi_{d\sigma}|\Psi_{a\sigma}>$, again same for the both spin components for a given TM.
\subsection{C\lowercase{ase-1,the adsorbate orbital is occupied, ${\large\varepsilon_a < \varepsilon_{d\sigma}}$}}
Suppose that the metal state is having the fractional occupancies respectively for the two  $f_\uparrow$ and $f_\downarrow$ for the two spin channels, 
In this case, the change in energy due to the adsorbate-metal interaction can be written as,
\begin{align}
\Delta E_d = & (\varepsilon_{b\uparrow}+\varepsilon_{b\downarrow}+f_\uparrow\varepsilon_{ab\uparrow}+
f_\downarrow\varepsilon_{ab\downarrow})-2\varepsilon_a-f_\uparrow\varepsilon_{d\uparrow}-f_\downarrow\varepsilon_{d\downarrow} \nonumber \\ &
=-(1-f_\uparrow)\frac{V^2}{\varepsilon_{d\uparrow}-\varepsilon_a}-(1-f_\downarrow)\frac{V^2}{\varepsilon_{d\downarrow}-\varepsilon_a}+(1+f_\uparrow)\alpha V^2+(1+f_\downarrow)\alpha V^2
\end{align}
Where $\alpha=-\frac{S}{V}$.
\subsection{C\lowercase{ase-2,the adsorbate orbital is unoccupied, ${\large \varepsilon_a > \varepsilon_{d\sigma}}$}}
The change in energy in this case is,
\begin{equation}
\Delta E_d =-f_{\uparrow}\frac{V^2}{\varepsilon_a-\varepsilon_{d\uparrow}}-f_\downarrow \frac{V^2}{\varepsilon_a-\varepsilon_{d\downarrow}}+f_\uparrow \alpha V^2+f_\downarrow \alpha V^2
\end{equation}
In the case of NH$_3$ molecule, we have 4 adsorbate orbitals, two among them are occupied(HOMO) and the remaining two are empty(LUMO), leading us to 
\begin{align}
\Delta E_d = & \underbrace{-\sum_{\sigma,i}f_\sigma\frac{V^2}{\varepsilon_i-\varepsilon_{d\sigma}}}_\text{d-LUMO interaction}
\underbrace{-\sum_{\sigma,j}(1-f_\sigma)\frac{V^2}{\varepsilon_{d\sigma}-\varepsilon_j}}_\text{d-HOMO interaction}
\underbrace{+\sum_{\sigma,i}f_\sigma\alpha V^2}_\text{d-LUMO orthogonalization} \nonumber \\ &
\underbrace{+\sum_{\sigma,j}(1+f_\sigma)\alpha V^2}_\text{d-HOMO orthogonalization}
\end{align}
Here the subscripts 'i' and 'j' respectively refers to LUMO and HOMO regions of $\varepsilon_a$.
\newpage
\vskip 0.2cm
\begin{figure}[h]
\includegraphics[width=120mm,height=120mm]{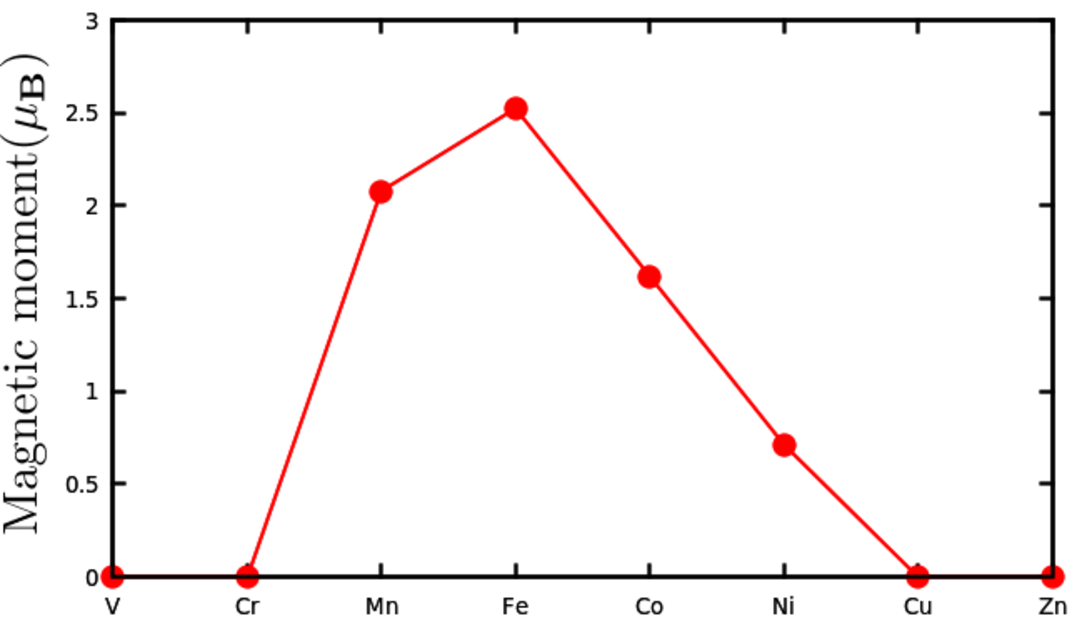}
\caption{(Color online) The magnetic moment of the TM atom attached to NH$_3$ for different TM-surfaces}
\label{Fig.2}
\end{figure}
\newpage
\section{O\lowercase{rigin of attractive and repulsive interactions: simple picture} }
To understand the interaction described by Eq.(2) of the manuscript,  we consider a simplified picture with a filled adsorbate state interacting with valence states of the TM-metal. Since there is no empty adsorbate state, the first and third term in the equation (2) are zero and we can write,
$\Delta E_d=-\sum_{\sigma,\sigma'}(1-f_\sigma)\frac{{V^{da}_{\sigma,\sigma'}}^2}{|\varepsilon_{d\sigma}-\varepsilon_{a\sigma'}|}+\sum_{\sigma,\sigma'}\alpha (1+f_\sigma){V^{da}_{\sigma,\sigma'}}^2
= (A_\uparrow +A_\downarrow)+(R_\uparrow +R_\downarrow)$. Where $A_\sigma$ and $R_\sigma$ are the spin dependent attractive and repulsive part of the metal-adsorbate interaction respectively.
\begin{equation}
\Delta E_d= (A_{\uparrow}+A_{\downarrow})+(R_\uparrow +R_\downarrow)
\end{equation}
where
$$A_\sigma=-\sum_{\sigma'}\frac{{V^{da}_{\sigma,\sigma'}}^2}{|\varepsilon_{d\sigma}-\varepsilon_{a\sigma'}|}
$$
is the energy gain due to formation of bonding orbitals for spin $\sigma$. Similarly,
$$
R_\sigma=\sum_{\sigma'}f_\sigma\frac{{V^{da}_{\sigma,\sigma'}}^2}{|\varepsilon_{d\sigma}-\varepsilon_{a\sigma'}|}+\sum_{\sigma'}\alpha (1+f_\sigma){V^{da}_{\sigma,\sigma'}}^2$$
is the repulsive energy due to the formation of antibonding orbitals for spin $\sigma$ \textit{plus} the energy due to the orthogonalization of adsorbate state to the metal.
Let us now consider an hypothetical case: the same TM can exist in two spin polarized states: (I) It is  100\% spin polarized, i,e spin-$\uparrow$ states are completely filled while spin-$\downarrow$ states are completely empty(half-metallic limit). (II) the TM is 0\% spin polarized (non-magnetic limit).
Let us consider the interaction of the filled adsorbate state with such states:
\subsection{Case I, half-metallic (HM) limit:} Let us consider that the TM is 100\% spin polarized i,e the majority spin states are completely filled and the minority spin states are completely empty. Such situation is represented in the  Fig.2a (top panel). In such case, the entire majority spin channel is pushed down so much that there is practically no attractive contribution from the majority spin ($A_\uparrow=0$). Similarly, there is no repulsive contribution from the minority spin ($R_\downarrow=0$), since we have assumed that the down-spin states are completely empty. From Eq. (9) we get,
\begin{equation}
\Delta E^{HM}_d= R_\uparrow+A_\downarrow,
\end{equation}
where $\Delta E^{HM}_d$ is the adsorption energy for the half-metallic case.\\

\subsection{\bf Case II, non-magnetic limit:} In this case, both spin channels contribute equally to the attractive and repulsive part (such situation in this case is represented in the Fig.2b (bottom panel)), and therefore , $A_\uparrow=A_\downarrow$ and $R_\uparrow=R_\downarrow$, From Eq. (10), we get
\begin{equation}
\Delta E^{NSP}_d= 2(R_\uparrow+A_\downarrow)=2\Delta E^{HM}_d,
\end{equation}
The adsorption energy for the non-magnetic case, $\Delta E^{NSP}_d$ is thus twice in magnitude compared to the adsorption energy in the half-metallic case. It is clear why a \textit{two-band centered d-band model} with appropriate filling is required to predict the catalytic activity on magnetically active surface. Calculation based on single centred model for spin polarized surfaces may lead over-binding(up to a factor of 2). 
\newpage 
\begin{figure}
\includegraphics[scale=0.5]{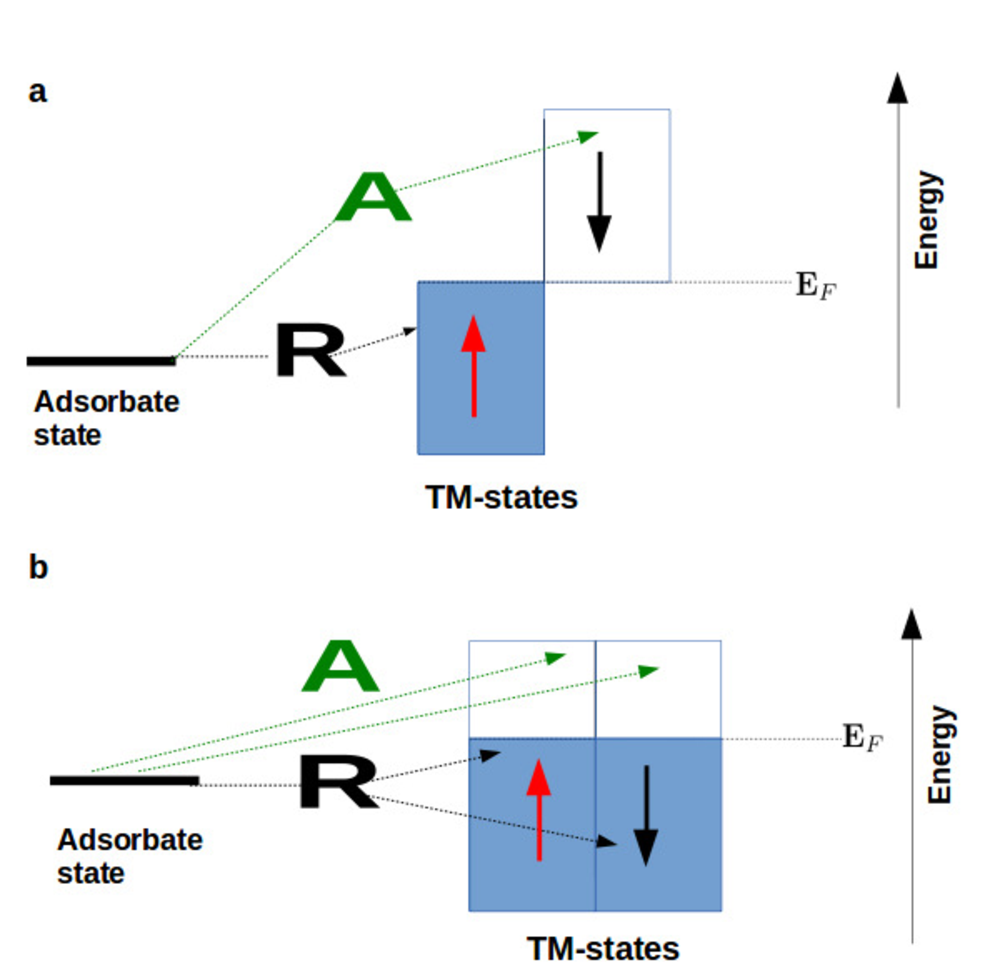}
\caption{(Color online) Interaction of a filled adsorbate state with the TM-states in two extream cases: (a)100\% spin polarization (half-metallic limit) (b) 0\% spin polarization (non-magnetic limit). Here "A" represents attractive interaction while "R" represent repulsive interaction between adsorbate-TM interactions.}
\label{Fig.2}
\end{figure}